\documentstyle[pre,aps,multicol]{revtex}

\tighten
\input epsf

\begin{document}
\draft

\title{\Large \bf  
Dynamical simulation of spin-glass and chiral-glass 
orderings \\ in 
three-dimensional Heisenberg spin glasses}

\author{\bf Hikaru Kawamura}

\address{Faculty of Engineering and Design, 
Kyoto Institute of Technology, Kyoto 606, Japan}

\maketitle

\begin{abstract}
Spin-glass and chiral-glass orderings in three-dimensional 
Heisenberg
spin glasses are
studied with and without randaom magnetic anisotropy
by dynamical Monte Carlo simulations.
In  isotropic case,
clear evidence of a finite-temperature
chiral-glass transition is presented. While the spin 
autocorrelation exhibits only
an interrupted aging, the chirality autocorrelation
persists to exhibit a pronounced aging
effect reminisecnt of the one observed
in the mean-field model. In 
anisotropic case, asymptotic mixing of the spin and the chirality
is observed in the off-equilibrium dynamics.
\end{abstract}


\begin{multicols}{2} \narrowtext
 
Recently, 
there arose a growing interest both theoretically and
experimentally in the off-equilibrium dynamical properties of 
glassy systems. In particular,
aging phenomena observed
in spin glasses [1]
have attracted  attention of researchers [2].
Unlike systems in thermal equilibrium,  relaxation of physical
quantites depends not only on the observation time $t$ but also on 
the waiting time $t_w$, {\it i.e.\/}, how long
one waits at a given state before the measurements. 
Recent studies  have revealed that
the off-equilibrium dynamics  in the spin-glass  state generally 
has two
characteristic time regimes [2,3]. One is a short-time 
regime, $t_0<<t<<t_w$ ($t_0$ is a microscopic time scale), 
called `quasi-equilibrium regime', and the
other is a long-time regime, $t>>t_w$, called `aging regime' or
`out-of-equilibrium regime'. In the quasi-equilibrium regime,
the relaxation is stationary and the fluctuation-dissipation
theorem (FDT) holds. The 
autocorrelation function at  times $t_w$ and $t+t_w$
is expected to behave as
\begin{equation}
C(t_w,t+t_w)\approx q^{{\rm EA}}+{C\over t^\lambda }\rightarrow
q^{{\rm EA}},
\label{qea}
\end{equation}
where $q^{{\rm EA}}$ is the equilibrium Edwards-Anderson order
parameter.
In the aging regime, the relaxation becomes non-stationary, 
FDT broken,
and the autocorrelation function decays to zero
as $t\rightarrow \infty $ for fixed 
$t_w$.

On theoretical side, both analytical and numerical studies 
of off-equilibrium dynamics of spin glasses have so far been
limited  to   {\it Ising-like\/} models, including the
Edwards-Anderson (EA) 
model with short-range interaction [4-6]
or the mean-field models with long-range interaction [3,7,8].  
Although these analyses on Ising-like  models
succeeded in reproducing some of the features of
experimental results, many of real spin-glass magnets are 
Heisenberg-like in the sense that the 
magnetic anisotropy is much
weaker than the isotropic exchange interaction.
Thus, in order to make a direct link between theory and experiment,
it is clearly desirable to study the dynamical
properties of {\it Heisenberg-like\/} spin-glass models.

Even at the static level,  nature of the experimentally
observed spin-glass
transtition and the spin-glass  state is not fully understood.
Although experiments have provided strong
evidence that spin-glass magnets exhibit 
an equilibrium  phase transition 
at a finite temperature, 
numerical studies  indicated
that the standard spin-glass order 
occurred only at zero temperature 
in  a three-dimensional (3D) Heisenberg spin glass [9-12]. 
While weak magnetic anisotropy inherent to real materials
is often invoked to explain this apparent discrepancy,
it remains puzzling that no
detectable sign of
Heisenberg-to-Ising crossover has been observed in experiments
which is usually expected to occur if the observed
spin-glass transition 
is caused by the weak magnetic anisotropy [9,10].

In order to solve this apparent puzzle, a  chirality mechanism of
experimentally observed spin-glass transitions
was recently proposed by the author [11],  on the assumption 
that an isotropic 3D Heisenberg
spin glass 
exhibited a finite-temperature 
{\it chiral-glass\/} transition without the conventional spin-glass 
order, in which only spin-reflection 
symmetry was broken with
preserving spin-rotation symmetry. `Chirality' is an Ising-like
multispin variable representing the
sense or the handedness of the noncollinear spin structures.
It was argued that, in real spin-glass magnets, 
the spin and the chirality were ``mixed'' due to the weak
magnetic ansitroy and the chiral-glass transition
was then ``revealed'' via anomaly in
experimentally accessible quantities. 
Meanwhile, 
theoretical question whether there really occurs such 
finite-temperature chiral-glass transition in an isotropic 3D
Heisenberg spin glass, 
remains somewhat inconclusive [11,12].

In view of the absence of off-equilibrium  simulation of
Heisenberg spin glasses, and also of the possible important role
played by the chirality,
I will report in the present Letter the results of  extensive
dynamical Monte Carlo simulations on  
isotropic and  anisotropic  3D Heisenberg
spin glasses, in which
the properties of both 
the spin and the chirality are studied. 

The model is the classical Heisenberg
model on 
a simple cubic lattice 
with the nearest-neighbor  random Gaussian couplings, $J_{ij}$ 
and $D_{ij}^{\mu \nu }$,
defined by the Hamiltonian
\begin{equation}
{\cal H}=-\sum_{<ij>} (J_{ij}{\mit\bf S}_i\cdot
{\mit\bf S}_j+D_{ij}^{\mu \nu }S_i^\mu S_i^\nu ),
\label{Hamil}
\end{equation}
where ${\bf S}_i$ 
=$(S_i^x,S_i^y,S_i^z)$ 
is a three-component 
unit vector, and 
the sum runs over all nearest-neighbor pairs 
with $N=L\times L\times L$ spins.   
$J_{ij}$ is the isotropic exchange coupling 
with zero mean and variance $J$,  while $D_{ij}^{\mu \nu}\ 
(\mu , \nu =x,y,z)$  is the random magnetic anisotropy 
with zero mean and variance $D$ which is assumed to be
symmetric and traceless,
$D_{ij}^{\mu \nu }=D_{ij}^{\nu \mu }$ and $\sum _\mu 
D_{ij}^{\mu \mu }=0$.

The local chirality at the $i$-th site and in the $\mu $-th 
direction, 
$\chi _{i\mu }$, may be defined for 
three neighboring spins 
by the scalar [9,12],
\begin{equation}
\chi _{i\mu }={\bf S}_{i+\hat {\bf e}_\mu }\cdot ({\bf S}_i\times 
{\bf S}
_{i-\hat {\bf e}_\mu }),
\end{equation}
where $\hat {\bf e}_\mu \ (\mu =x,y,z)$ 
denotes a unit lattice vector along the
$\mu\/$-axis.
Note that the chirality defined by Eq.(3) is a pseudoscalar in the
sense that it
is invariant under global spin rotation but changes sign under 
global spin reflection or inversion. 

The spin and  chirality
autocorrelation functions are
defined by
\begin{equation}
$$C_s(t_w,t+t_w)={1\over N}
\sum _i[<\vec S_i(t_w)\cdot \vec S_i(t+t_w)>],
\end{equation}
\begin{equation}
C_\chi (t_w,t+t_w)={1\over 3N}\sum _{i,\mu }[<\chi _{i\mu }(t_w) 
\chi _{i\mu }(t+t_w)>],
\end{equation}
where $<\cdots >$ represents the thermal average and $[\cdots ]$ 
represents the average over bond disorder.

Monte Carlo simulation is performed based on the standard single
spin-flip heat-bath method.
Starting from completely random initial
configurations, the system is  
quenched to a working temperature.
Total of about $3\times 10^5$ 
Monte Carlo steps per spin [MCS] are generated
in each run.
Sample average is taken over 30-120 independent bond realizations.
The lattice size  mainly studied is $L=16$ with periodic
boundary conditions, while in some cases lattices with  $L=12$
and $24$ are also studied.
\par\smallskip

Let us begin with the fully isotropic case, $D=0$.
The spin and chirality autocorrelation functions 
at a low temperature $T/J=0.05$ are
shown in Fig.1 as a function of $t$.
For larger $t_w$,
the curves of the spin autocorrelation function  $C_s$ 
come on top of each other
in the long-time
regime, indicating that the stationary
relaxation is recovered and
aging  is
interrupted. This behavior has been expected because the 3D
Heisenberg spin glass is believed to have 
no standard spin-glass order [9-12]. 
Similar interrupted aging was observed 
in the 2D Ising
spin glass which did not have an equilibrium spin-glass order
[5]. 
By contrast, the chiral 
autocorrelation function $C_{\chi }$ 
shows an entirely different behavior:
Follwoing the initial decay, it exhibits a  clear plateau
around $t\sim t_w$ and then drops sharply for $t>t_w$. 
It also shows an eminent aging effect, namely, as one waits longer,
the relaxation becomes slower and the plateau-like behavior at
$t\sim t_w$ becomes more pronounced. 

\smallskip\centerline{
\epsfysize=9cm
\epsfxsize=6.5cm
\epsfbox{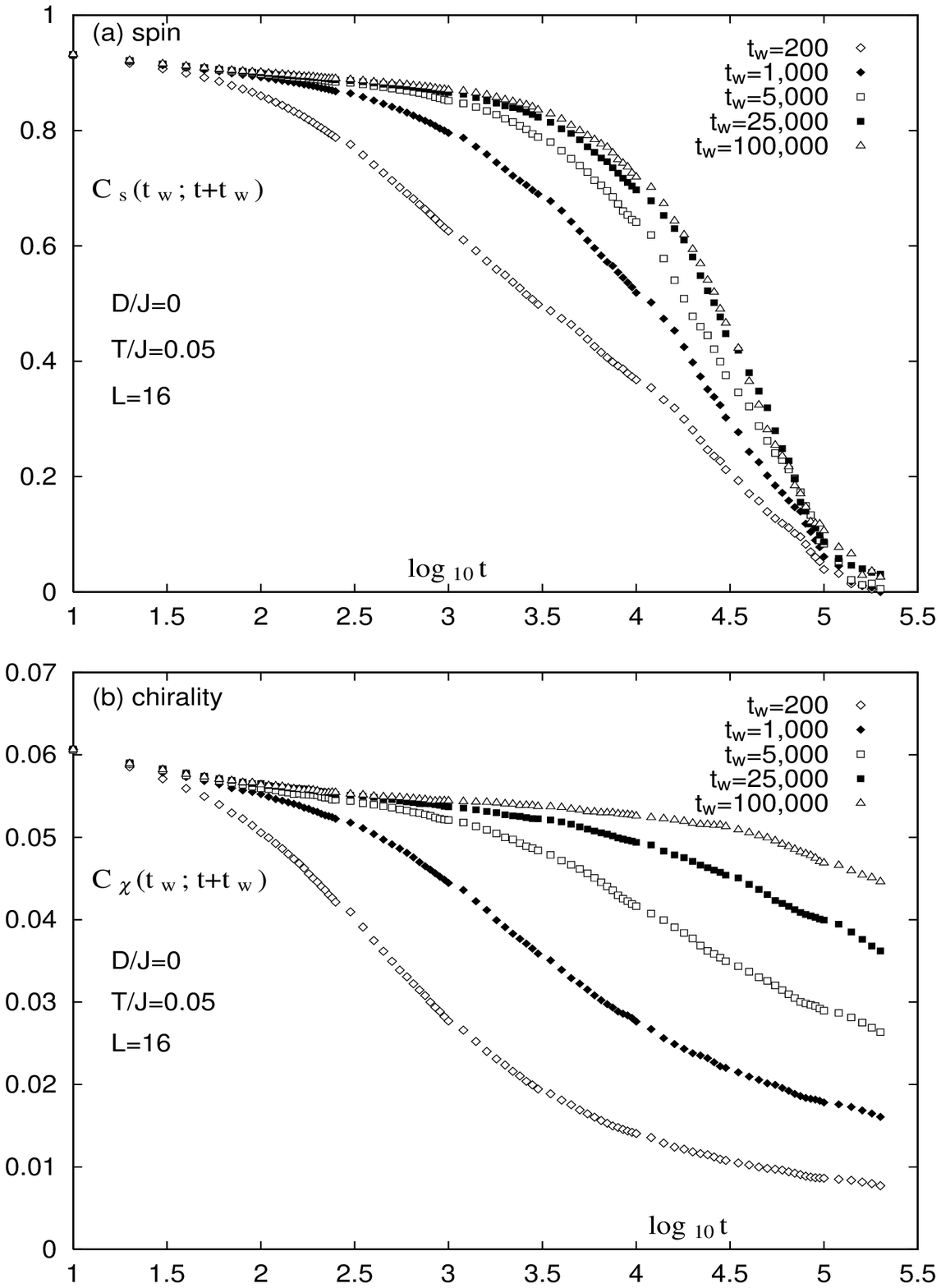}}
\noindent
FIG.1~ Spin (a) and chirality (b) autocorrelation functions
of a 3D  isotropic Heisenberg spin glass at a temperature $T/J=0.05$
plotted versus 
$\log _{10}t$ for various waiting times $t_w$. The lattice size
is $L=16$ averaged over 66 samples.

\bigskip\centerline{
\epsfysize=9cm
\epsfxsize=6.5cm
\epsfbox{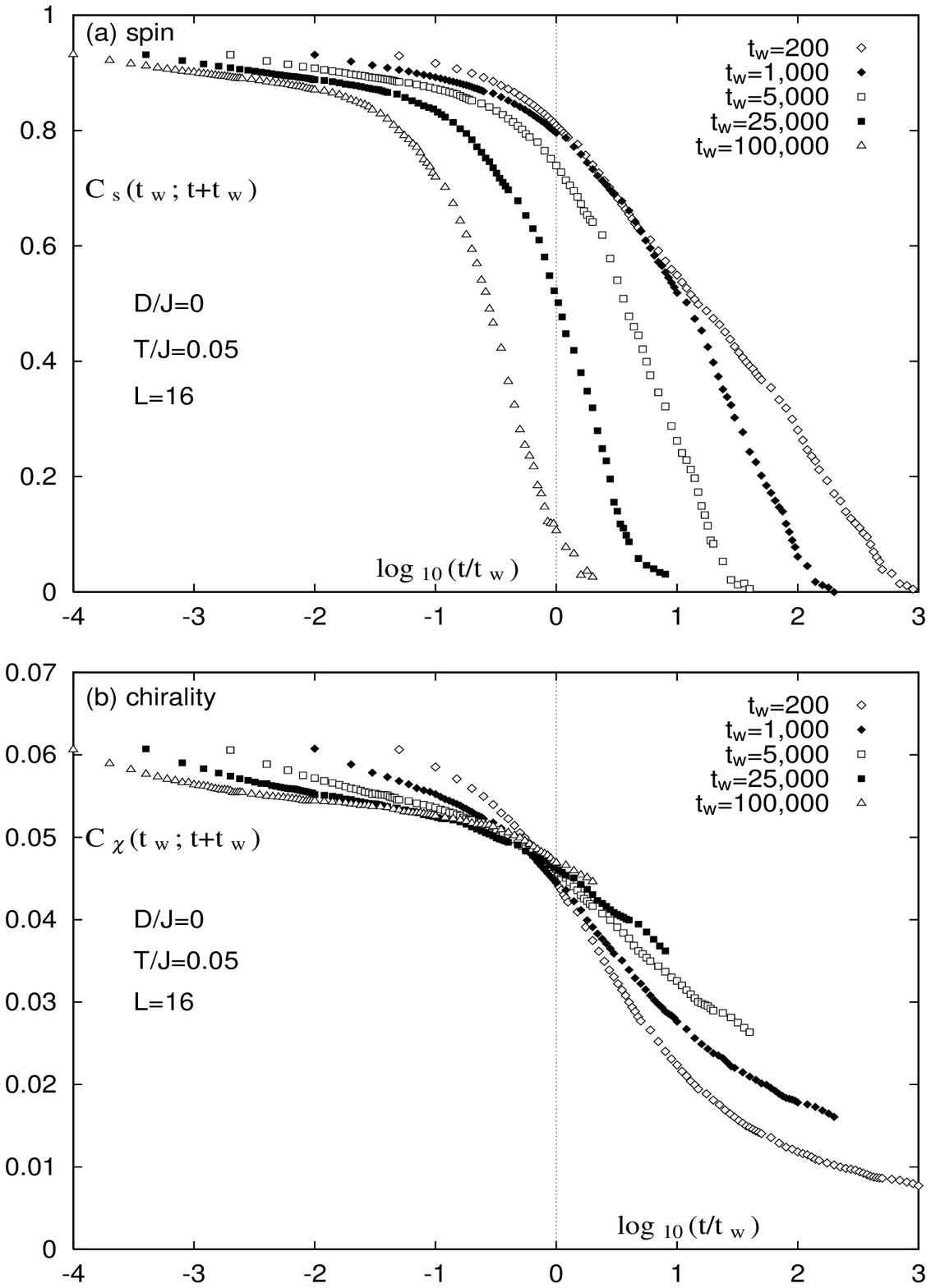}}
\noindent
FIG.2~ The same data as in Fig1, 
but plotted versus $\log _{10}(t/t_w)$.

\bigskip
In Fig.2, $C_s$ and $C_\chi $  are replotted
as a function of the
scaled time $t/t_w$.
Reflecting its interrupted aging, the curves of $C_s$ 
for larger $t_w$ now lie below the ones
for smaller $t_w$
({\it subaging\/}). 
By contrast, the curves of $C_\chi $ for
various $t_w$ cross around $t/t_w\sim 1$, and at $t>t_w$, 
the data for larger $t_w$ lie {\it above\/} the ones 
for smaller $t_w$ ({\it superaging\/}).
Such superaging behavior of $C_\chi $ means 
that the aging in chirality is
more enhanced than the one expected from the naive 
$t/t_w$-scaling. Note that, although the
chirality is an Ising-like variable from  symmetry,
the observed superaging behavior is in  contrast to the aging
behavior of the 3D EA model
which was found 
to satisfy a
good $t/t_w$-scaling in the aging regime [5]. 
It should also be noticed that 
the plateau-like behavior observed here has been 
hardly noticeable
in  simulations of the 3D EA model.
Rather, the behavior of $C_\chi $ observed here
is reminiscent of the one observed in
the mean-field model
such as the Sherrington-Kirkpatrick (SK) 
model [7,8]. This correspondence might suggest
that an effective interaction between the chiralities is long-ranged.

While the 
plateau-like behavior observed in $C_\chi $ is already suggestive 
of a nonezero {\it chiral\/} Edwards-Anderson order parameter, 
$q_{{\rm CG}}^{{\rm EA}}>0$,  more quantitative
analysis similar to the one recently done by Parisi {\it et al\/}
for the 4D Ising spin glass [6] is performed  
to extract $q_{{\rm CG}}^{{\rm EA}}$ from the data of
$C_\chi $ in the quasi-equilibrium regime.
Finiteness of $q_{{\rm CG}}^{{\rm EA}}$ is also visible in a
log-log plot of
$C_\chi $ versus $t$ as shown in the inset 
of Fig.3, where the data show a clear upward curvature.
I extract $q_{{\rm CG}}^{{\rm EA}}$ by fitting 
the data of $C_\chi $ for
$t_w=3\times 10^5$
to the power-law form of eq.(1) in the time range
$40\leq t\leq 3,000$.
The obtained $q_{{\rm CG}}^{{\rm EA}}$,
plotted as a functon of temperature in Fig.3,  
clearly indicates the
occurrence of a finite-temperature chiral-glass transition
at $T_{{\rm CG}}/J=0.157\pm 0.01$ with the associated 
order-parameter exponent
$\beta _{{\rm CG}}=1.1\pm 0.1$. 
The size dependence turns out to be rather small,
although the mean values of $q_{{\rm CG}}^{{\rm EA}}$
tend to slightlty
increase around $T_{{\rm CG}}$ with increaing $L$.
Since both  finite-size effect and finite-$t_w$ effect
tend to underestimate  $q_{{\rm CG}}^{{\rm EA}}$, one may
regard the present result as a rather strong
evidence of the occurrence of a finite-temperature 
chiral-glass transition. 

\bigskip\centerline{
\epsfysize=5cm
\epsfxsize=7cm
\epsfbox{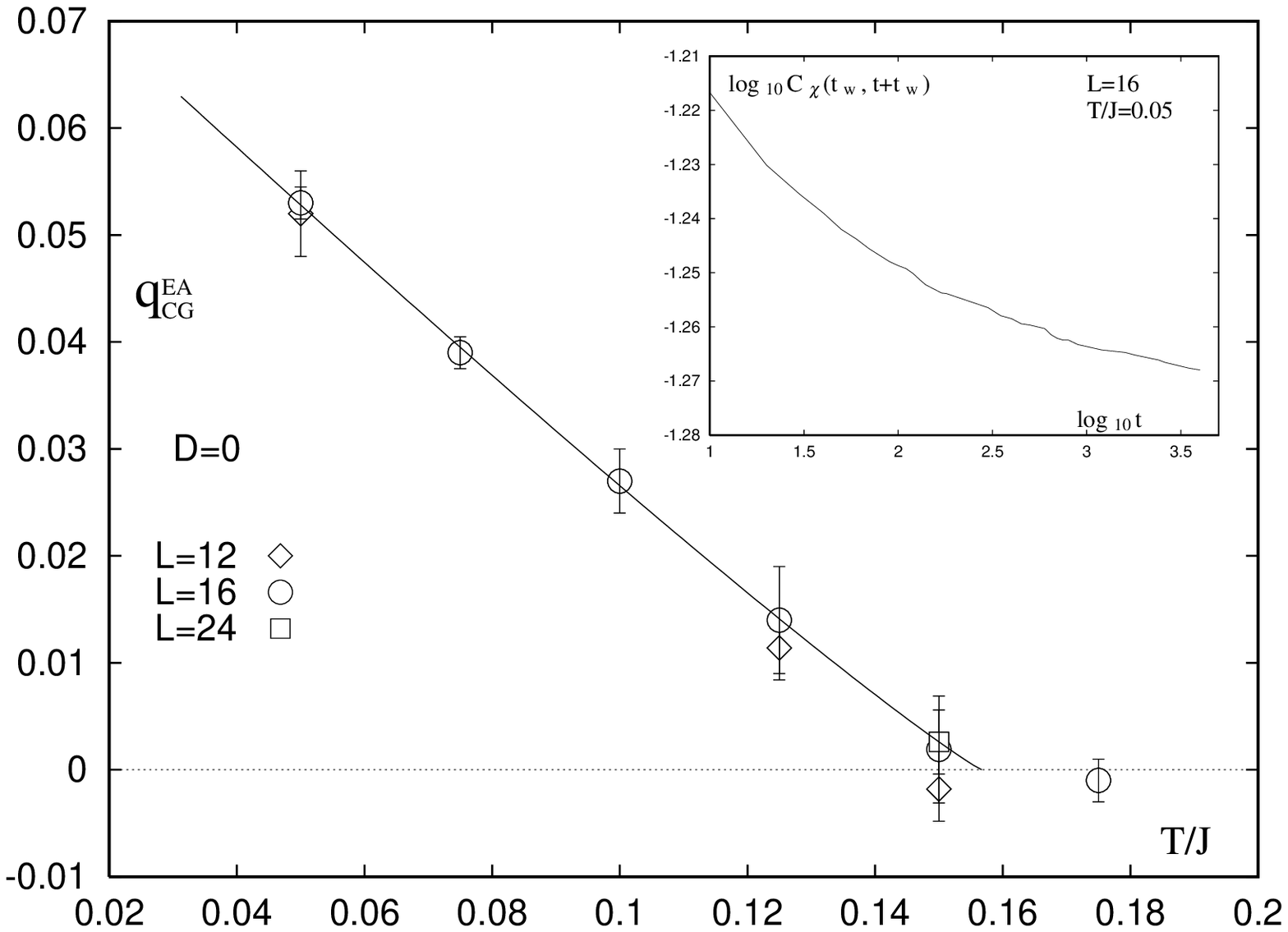}}
\noindent
FIG.3~ Temperature dependence of the Edwards-Anderson order
parameter of the chirality of a 3D  isotropic Heisenberg spin glass.
The data are averaged over 30-120 samples.
Inset exhibits the log-log plot of the
$t$-dependence of the
chiral autocorrelation function in the quasi-equilibrium regime
for $L=16$, $T/J=0.05$ and
$t_w=3\times 10^5$.

\bigskip
The associated exponent $\beta _{{\rm CG}}\sim 1.1$ 
is considerably
larger than the value of the 3D EA model $\beta \sim 0.5$ [10], 
but is close to the value of the mean-field model $\beta =1$.
This suggests that the universality class of the chira-glass
transition of the 3D Heisenberg spin glass might be 
different from that of the standard 3D Ising spin glass.
According to the chirality mechanism,
the criticality of real
spin-glass transitions 
should be the same as that of the chiral-glass
transition of an isotropic Heisenberg spin glass, so long as the
magnitude of  random anisotropy is not too strong.
If one tentatively accepts this scenario, 
the present result opens up a new interesting possiblity
that the universality class of many of real spin-glass transitions 
might differ from that of the standard Ising
spin glass, contrary to common belief.

\bigskip
In the presence of
weak  anisotropy $D>0$, chirality scenario predicts at the static level that
the transition behavior of  chirality remains essentially 
the same
as in the isotropic case, whereas the spin is 
mixed into the chirality,
asymptotically
showing the same transition behavior as the chirality [11].
In order to see whether such ``spin-chirality mixing'' occurs  in 
the off-equilibrium dynamics, 
further dynamical simulations are performed for the models with 
random anisotropies $D/J=0.01\sim 1$.
While
chirality exhibits essentially the same dynamical behavior
as in the isotropic case (not shown here),
the  behavior
of spin at $t>t_w$ 
changed significantly in the 
presence
of  anisotropy. 
As an example, the spin autocorrelation in the case of 
weak anisotropy $D/J=0.01$ is shown in Fig.4.
Even for such small anisotropy,
spin is found to show
{\it superaging\/} behavior asymptotically
at $t>>t_w$ similar to that
of the chirality in the fully isotropic case, 
demonstrating the spin-chirality mixing.

\bigskip\centerline{
\epsfysize=5cm
\epsfxsize=7cm
\epsfbox{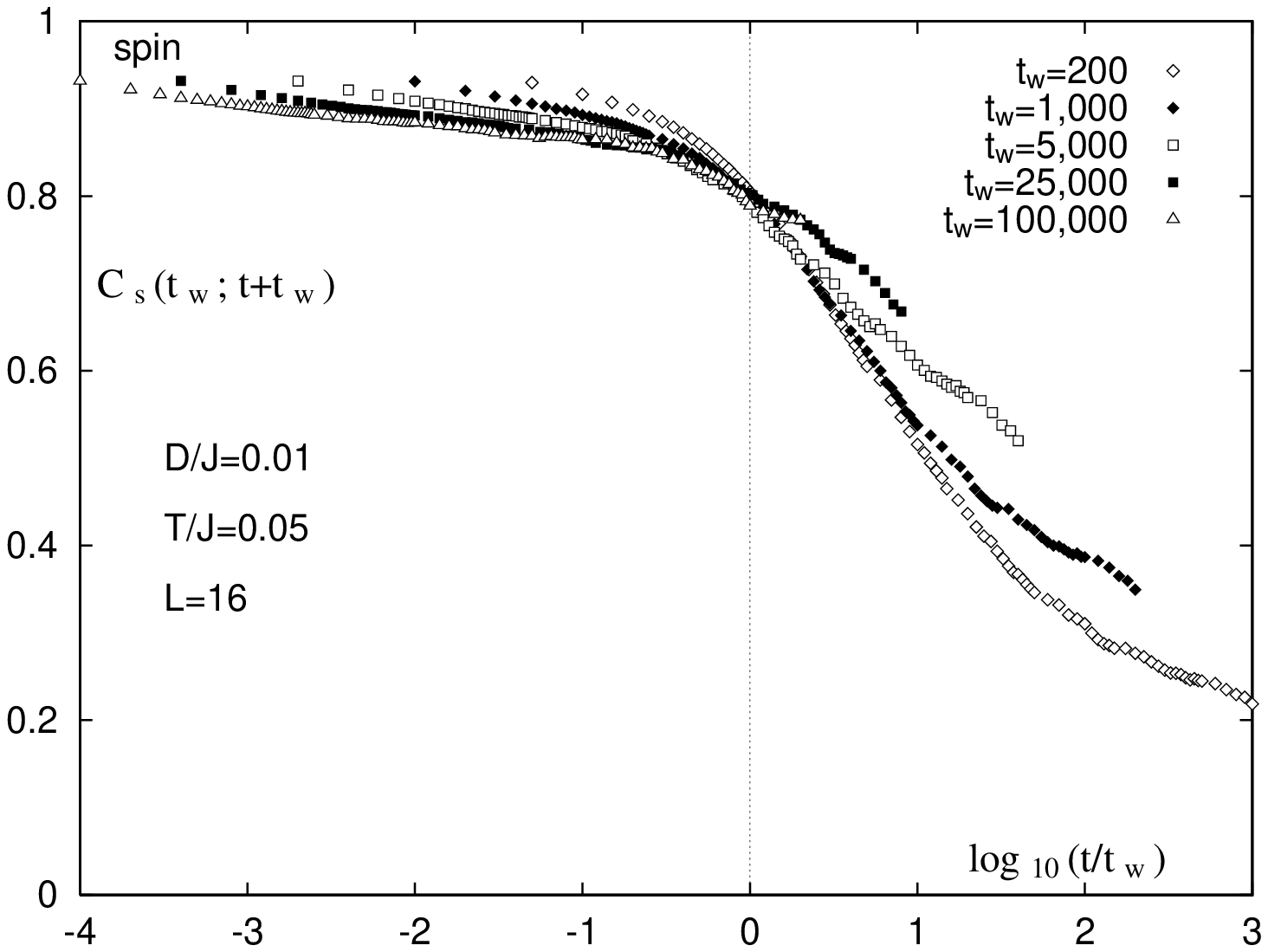}}
FIG.4~  Spin autocorrelation function
of the weakly anisotropic 3D Heisenberg spin glass
with $D/J=0.01$
plottedd versus
$\log _{10}(t/t_w)$.
The lattice size is
$L=16$ averaged over 60 samples and the temperature is $T/J=0.05$.

\bigskip
Experimentally,  thermoremanent magnetization (TRM) or 
zero-field-cooled (ZFC) magnetization is found to
show an approximate $t/t_w$-scaling
in the aging regime, with  small deviation from
the perfect scaling
in the direction of subaging [2]. Although this seems 
in apparent contrast to the present result,
it should be noticed that standard aging experiments 
have  been made by measuring the magnetic response,
not the autocrrelation. 
Recent numerical simulation by Yoshino {\it et al\/}  
revealed that, at least in the case of the 
SK model, 
TRM showed the subaging 
even when the spin correlation showed the superaging [13].
Thus, I also calculate the ZFC magnetization 
for an anisotropic model
with $D/J=0.05$:
After the initial quench, 
the system is  evolved in zero field during $t_w$ MCS. 
Then, an external field
of instensity $H/J=0.05$ is turned on and the subsequent growth 
of the magnetization
$M(t;t_w)$ is recorded. As can be seen from Fig.5, 
the data show the near $t/t_w$-scaling
in the aging regime $t>t_w$ where the 
spin-autocorrelation shows the superaging.
Thus, the observed tendency is roughly
consistent with experiments.  
It might be interesting to experimentally
investigate the aging properties of {\it spin correlations\/}
of Heisenberg-like magnets in search for  possible
superaging behavior.

\bigskip\centerline{
\epsfysize=5cm
\epsfxsize=7cm
\epsfbox{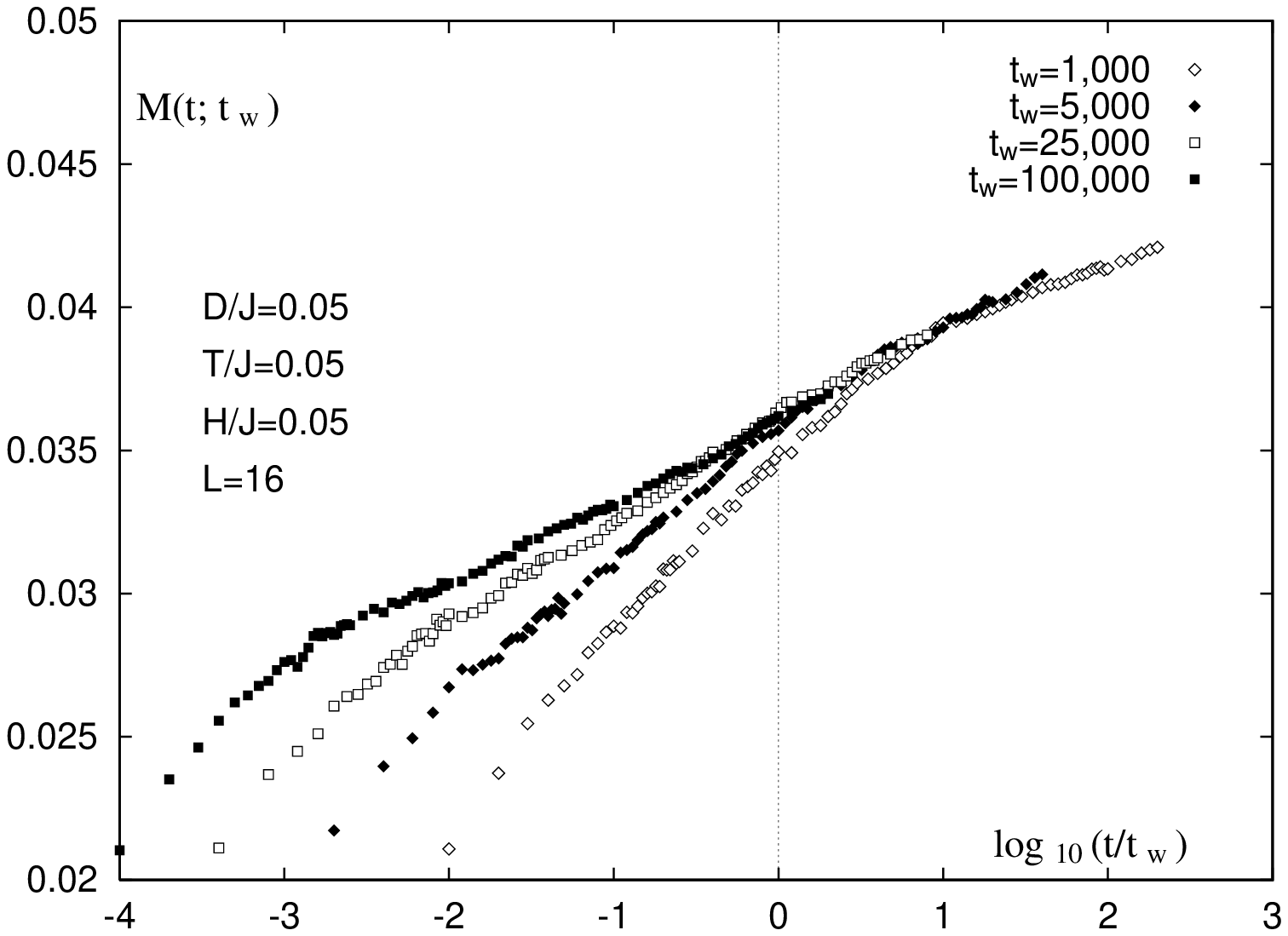}}
\noindent
FIG.5~  Zero-field-cooled  magnetization
of an   anisotropic 3D Heisenberg spin glass with $D/J=0.05$
plotted versus $\log _{10}t$.
The field is  $h/J=0.05$ and
the temperature is $T/J=0.05$.
The lattice size is  $L=16$
averaged over 80 samples.

\bigskip
In summary, equilibrium and off-equilibrium properties of the spin and chirality order
in 3D Heisenberg spin glasses are studied 
with and without random anisotropy 
by dynamical Monte Carlo simulations. 	
The results are basically consistent with the chirality
mechanism: 
In the isotropic case,  spin and chirality
show very different dynamical behaviors consistent with the
`spin-chirality separation', whereas in the anisotropic case, 
spin shows the same asymptotic behavior as  
chirality, consistent
with the `spin-chirality mixing' due to magnetic  anisotropy.
Furthermore, clear evidence for the occurrence of a 
finite-temperature chiral-glass transition in an isotropic 3D
Heisenberg spin glass is  presented.

The author is thankful to  H.Takayama, E.Vincent, 
L.F.Cugliandolo, 
M.Ocio,  H.Rieger, K.Hukushima and H.Yoshino for useful
discussion. The numerical calculation was performed on the FACOM
VPP500 at the supercomputer center, ISSP, University of Tokyo.

\end{multicols}\widetext

\end{document}